\documentclass{article}
\usepackage{spconf,amsmath,graphicx}

\usepackage{enumitem}
\usepackage{bbm}
\setlist{nosep, leftmargin=14pt}
\usepackage{amsmath}
\usepackage{amssymb}
\usepackage{mwe} 

\newcommand{\Dice}{\mathcal{D}}
\newcommand{\CE}{\mathcal{H}}
\newcommand{\BCE}{\mathcal{B}}

\newcommand{\fEMA}{f^{\text{ema}}}
\newcommand{\Mix}{\operatorname{Mix}}

\usepackage[]{acronym}
\acrodef{DL}{Deep Learning}
\acrodef{US}{Ultrasound}
\acrodef{PCCL}{{\bf P}ixel-level and {\bf C}lass-level {\bf C}onsistency {\bf L}earning}
\acrodef{CNN}{Convolutional Neural Network}
\acrodef{ROI}{Region of Interest}
\acrodef{SSL}{semi-supervised learning}
\acrodef{GT}{Ground Truth}
\acrodef{CR}{Consistency Regularization}
\acrodef{EMA}{Exponential Moving Average}
\acrodef{MAC}{Mutual Agreement Consistency}
\acrodef{KL}{Kullback–Leibler divergence}
\acrodef{MIG}{Mutual Information Gap}
\acrodef{CE}{Cross-Entropy}
\acrodef{Dice}{Dice based coefficient}
\acrodef{FM-DACL}{Foundation Model Dual Agreement Consistency Learning}

\title{Dual Agreement Consistency Learning with Foundation Models for Semi-Supervised Fetal Heart Ultrasound Segmentation and Diagnosis}
\name{Fangyijie Wang$^{\star \ddagger \circ}$\thanks{$\star$ Corresponding author (fangyijie.wang@ucdconnect.ie)} \thanks{This work was funded by Taighde \'{E}ireann – Research Ireland through the Centre for Research Training in Machine Learning (18/CRT/6183).} \qquad Gu\'enol\'e Silvestre$^{\ddagger \dagger}$ \qquad Kathleen M. Curran$^{\ddagger \circ}$}
\address{$^{\ddagger}$ Taighde \'{E}ireann – Research Ireland Centre for Research Training in Machine Learning, Ireland \\
$^{\dagger}$ School of Computer Science, University College Dublin, Ireland \\
$^{\circ}$ School of Medicine, University College Dublin, Ireland \\
}
\begin{document}
%
\maketitle

\begin{abstract}

\end{abstract}

Congenital heart disease (CHD) screening from fetal echocardiography requires accurate analysis of multiple standard cardiac views, yet developing reliable artificial intelligence models remains challenging due to limited annotations and variable image quality. In this work, we propose FM-DACL, a semi-supervised Dual Agreement Consistency Learning framework for the FETUS 2026 challenge on fetal heart ultrasound segmentation and diagnosis. The method combines a pretrained ultrasound foundation model (EchoCare) with a convolutional network through heterogeneous co-training and an exponential moving average teacher to better exploit unlabeled data. Experiments on the multi-center challenge dataset show that FM-DACL achieves a Dice score of 59.66 and NSD of 42.82 using heterogeneous backbones, demonstrating the feasibility of the proposed semi-supervised framework. These results suggest that FM-DACL provides a flexible approach for leveraging heterogeneous models in low-annotation fetal cardiac ultrasound analysis. The code is available on https://github.com/13204942/FM-DACL.

\section{Introduction}

Congenital heart disease (CHD) is the most common fetal structural anomaly and a leading cause of neonatal morbidity and mortality. Prenatal screening relies on ultrasound (US) examination of multiple standard cardiac planes, including the four-chamber heart (4CH), left ventricular outflow tract (LVOT), right ventricular outflow tract (RVOT), and three-vessel-and-trachea (3VT) views, which together improve diagnostic sensitivity compared with using the 4CH view alone \cite{Bravovalenzuela:2018,Ogge:2006}. However, accurate interpretation of these views requires substantial clinical expertise and remains subject to inter-observer variability. In addition, fetal cardiac structures often present subtle anatomical patterns and variable image quality, making reliable analysis challenging \cite{Wu:2009}. Although artificial intelligence has shown promise in assisting prenatal ultrasound analysis, progress is limited by the scarcity of large, well-annotated datasets and the high cost of expert labeling \cite{DAlberti:2025}.

The FETUS 2026 challenge addresses these limitations by providing a large-scale, multi-center fetal echocardiography dataset containing 5,000 images collected from seven hospitals and multiple ultrasound systems. The challenge focuses on a semi-supervised setting in which only 10\% of the training data are annotated with both segmentation masks and diagnostic labels. Participants are required to develop models that simultaneously perform cardiac structure segmentation and CHD classification from a single ultrasound image, encouraging robust and efficient AI solutions for fetal cardiac assessment.

To address the \ac{SSL} setting of this challenge, we propose \ac{FM-DACL}, a unified framework that leverages heterogeneous models to effectively utilize limited labeled data and abundant unlabeled data. The proposed method combines a pretrained ultrasound foundation model, EchoCare, with a lightweight U-Net as a complementary network. These models are jointly optimized using supervised losses on labeled samples and dual agreement consistency on unlabeled data. Specifically, \ac{FM-DACL} enforces prediction consistency at both local (pixel-wise segmentation) and global (classification-level) representations, encouraging stable segmentation outputs and consistent diagnostic predictions. By exploiting complementary inductive biases between the foundation model and the convolutional network, the proposed approach improves robustness and generalization under limited annotation conditions.

\section{Methodology}

\begin{figure}
\begin{center}
\includegraphics[width=\linewidth]{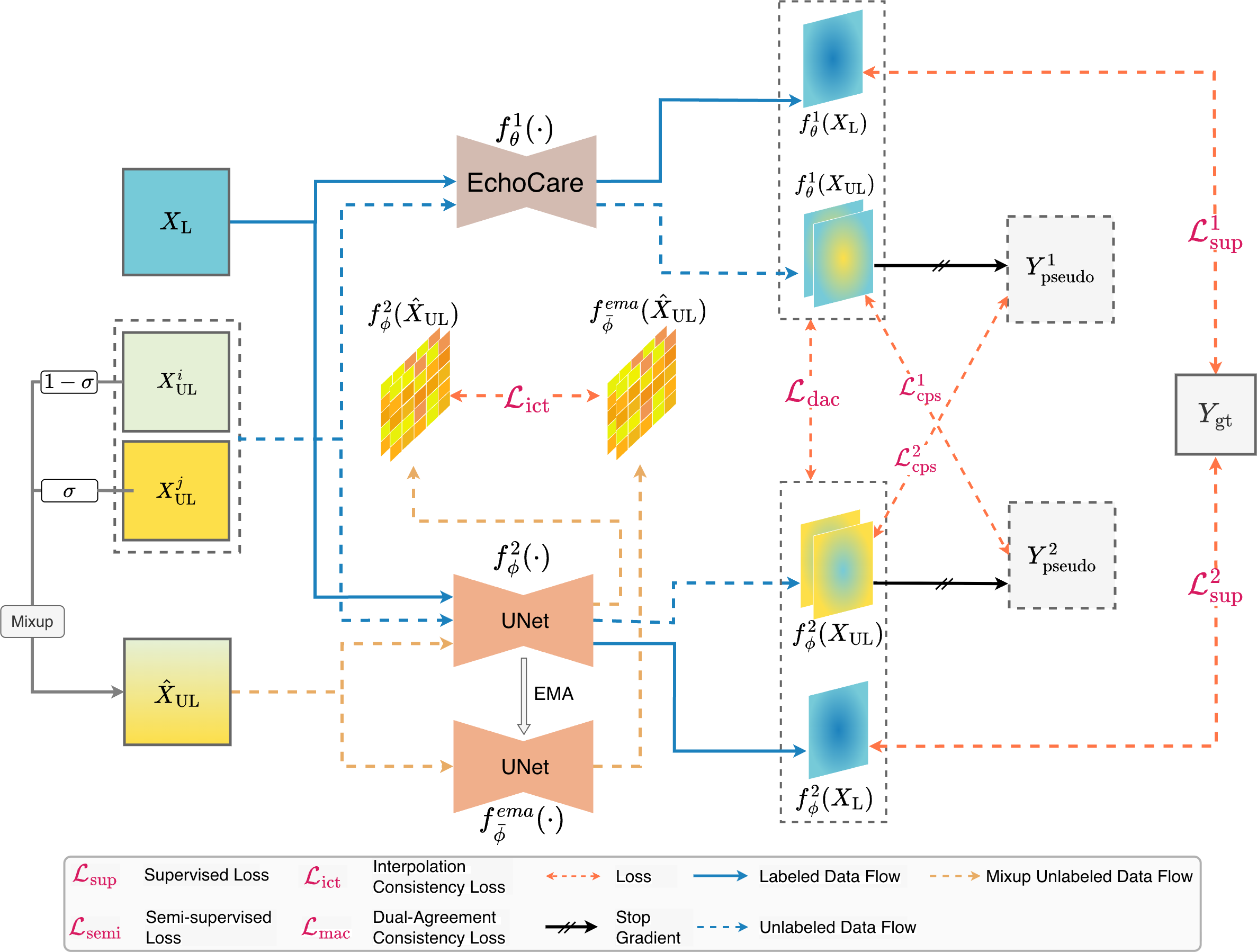}
\end{center}
   \caption{An overview of our \ac{FM-DACL} method with EchoCare and U-Net for semi-supervised image segmentation.}
\label{fig:framework}
\end{figure}

This section reviews EchoCare and U-Net. Subsequently, we explain our semi-supervised framework \ac{FM-DACL} in the following sections~\ref{ssec:cotrain},~\ref{ssec:ict},~\ref{ssec:dac}, and~\ref{ssec:learningloss}. The overview of our method is illustrated in Fig.\ref{fig:framework}.

\subsection{Preliminaries}

The proposed \ac{FM-DACL} adopts a heterogeneous architecture composed of a foundation model and a convolutional network. Specifically, we use the pretrained EchoCare \cite{zhang:2025} foundation model as the primary network $f^1_{\theta}(\cdot)$, which provides strong representation capability through large-scale pretraining on echocardiography data. EchoCare serves as a high-capacity encoder for extracting rich anatomical features from fetal ultrasound images \cite{zhang:2025}.
To complement the foundation model, we employ a standard U-Net \cite{Ronneberger:2015} architecture as the network $f^2_{\phi}(\cdot)$. U-Net provides an efficient convolutional encoder--decoder structure that captures fine-grained spatial details while maintaining computational efficiency \cite{Ronneberger:2015}. The two networks exhibit complementary inductive biases: the foundation model offers powerful global feature representations, whereas the convolutional network preserves detailed local structures.
This heterogeneous design forms the basis of the proposed \ac{FM-DACL} framework. During training, knowledge is exchanged between the two networks through dual agreement consistency, enabling effective \ac{SSL} from limited labeled data while leveraging the large amount of unlabeled samples available in the dataset.

\subsection{Co-training between CNN and Transformer}
\label{ssec:cotrain}

\ac{FM-DACL} adopts a cross-supervision strategy between the two heterogeneous networks: the EchoCare foundation model $f^1_{\theta}(\cdot)$ and the U-Net $f^2_{\phi}(\cdot)$. This mechanism encourages the two models to learn from each other's predictions, allowing complementary knowledge transfer during training.

Given an unlabeled input batch $\boldsymbol{X}_{\text{UL}}$, the two networks independently produce prediction maps $f^1_{\theta}(\boldsymbol{X}_{\text{UL}})$ and $f^2_{\phi}(\boldsymbol{X}_{\text{UL}})$. Pseudo labels are then generated by converting the predicted probability maps into one-hot representations:
\[
\tilde{Y}^{1} = f_{\text{OH}}\big(f^1_{\theta}(\boldsymbol{X}_{\text{UL}})\big), 
\quad
\tilde{Y}^{2} = f_{\text{OH}}\big(f^2_{\phi}(\boldsymbol{X}_{\text{UL}})\big),
\]
where $f_{\text{OH}}(\boldsymbol{x}=c)=\mathbbm{1}_{[\boldsymbol{x}=c]}$ denotes the one-hot encoding operator.
During optimization, the pseudo labels produced by one network are used to supervise the other network. Specifically, $\tilde{Y}^{1}$ supervises $f^2_{\phi}$ and $\tilde{Y}^{2}$ supervises $f^1_{\theta}$, while gradients are not propagated between predictions and their own pseudo labels. This cross-supervision mechanism allows both networks to iteratively refine their predictions on unlabeled data and improves the robustness of the \ac{SSL} process.

The cross-supervision objective is formulated as a semi-supervised loss $\mathcal{L}_{\text{cps}}$, composed of two reciprocal terms $\mathcal{L}_{\text{cps}}^{1}$ and $\mathcal{L}_{\text{cps}}^{2}$. Each network is trained using pseudo labels generated by the other network for both segmentation and classification tasks. The loss is defined as
{\small
\[
\label{cps}
\begin{aligned}
\mathcal{L}_{cps} =
\sum_{i=1}^{2}
& [\mathcal{H}\big(f_i(\boldsymbol{X}_{UL}), \tilde{\boldsymbol{Y}}_{j}\big)
+ \mathcal{D}\big(f_i(\boldsymbol{X}_{UL}), \tilde{\boldsymbol{Y}}_{j}\big) \\
&+ \mathcal{B}\big(g_i(\boldsymbol{X}_{UL}), \tilde{\boldsymbol{C}}_{j}\big) ], \; i \neq j
\end{aligned}
\]
}
where $i,j \in \{1,2\}$ and $i \neq j$, indicating that each network is supervised by pseudo-labels generated from the other network. $\CE(\cdot)$ and $\Dice(\cdot)$ denote the cross-entropy and Dice losses for the segmentation task. $\BCE(\cdot)$ denotes the binary cross-entropy with logits loss used for multi-label classification. $f^1_{\theta}(\cdot)$ and $f^2_{\phi}(\cdot)$ represent the segmentation predictions of EchoCare and U-Net, respectively, while $g^1_{\theta}(\cdot)$ and $g^2_{\phi}(\cdot)$ denote their corresponding classification outputs. $\tilde{Y}$ and $\tilde{C}$ represent pseudo labels for segmentation and classification generated from the counterpart network. During inference, only the EchoCare foundation model (i.e., $f^1_{\theta}$) is used to produce the final segmentation and classification predictions.

\subsection{Interpolation Consistency Learning}
\label{ssec:ict}

\ac{FM-DACL} uses an interpolation consistency learning strategy for the network $f^2_{\phi}(\cdot)$. This strategy encourages the model to produce consistent predictions when inputs are linearly interpolated, thereby improving generalization and robustness.

Following the mean-teacher paradigm, we maintain a teacher model $\fEMA_{\bar{\phi}}(\cdot)$ whose parameters $\bar{\phi}$ are updated as the exponential moving average (EMA) of the student parameters $\phi$. Given two unlabeled images $\boldsymbol{X}^i_{\text{UL}}$ and $\boldsymbol{X}^j_{\text{UL}}$, we generate an interpolated sample
$
\hat{\boldsymbol{X}}_{\text{UL}} = \Mix(\boldsymbol{X}^i_{\text{UL}}, \boldsymbol{X}^j_{\text{UL}})
$,
where $\Mix(\cdot)$ denotes the mixup operation with combination ratio $\sigma = 0.5$, and $i+j$ equals the unlabeled batch size $\text{B}_{\text{UL}}$.

The student network prediction for the mixed input $f^2_{\phi}(\hat{\boldsymbol{X}}_{\text{UL}})$ is encouraged to match the interpolation of the corresponding teacher predictions $\Mix(\fEMA_{\bar{\phi}}(\boldsymbol{X}^i_{\text{UL}}), \fEMA_{\bar{\phi}}(\boldsymbol{X}^j_{\text{UL}}))$. The interpolation consistency loss is defined as
{\small
\[
\mathcal{L}_{\text{ict}} =
\sum \left(
f^2_{\phi}(\hat{\boldsymbol{X}}_{\text{UL}}) -
\Mix\big(
\fEMA_{\bar{\phi}}(\boldsymbol{X}^i_{\text{UL}}),
\fEMA_{\bar{\phi}}(\boldsymbol{X}^j_{\text{UL}})
\big)
\right)^2.
\]
}
This interpolation consistency regularization encourages smooth prediction behavior in the input space and enables more effective learning from unlabeled ultrasound images.

\subsection{Dual-Agreement Consistency}
\label{ssec:dac}

We introduce a dual agreement consistency objective that promotes collaborative learning between the EchoCare foundation model $f^1_{\theta}(\cdot)$ and the U-Net model $f^2_{\phi}(\cdot)$. The key idea is to encourage both networks to produce compatible predictions on unlabeled images while maintaining confident and stable outputs.

The proposed agreement regularization contains two complementary terms. First, a pixel-wise distribution alignment loss encourages the predicted class probabilities from the two networks to be consistent. Second, an entropy-based agreement term promotes confident and mutually compatible predictions by penalizing disagreement in uncertainty between the two models. Together, these components encourage the models to exchange information during training while mitigating noisy pseudo-supervision. Formally, the dual agreement consistency loss is defined as
$
\mathcal{L}_{\mathrm{dac}} = \mathcal{L}_{\mathrm{align}} + \mathcal{L}_{\mathrm{conf}},
$
where the two components are computed as
{\small
\[
\begin{aligned}
\mathcal{L}_{\mathrm{align}} &= 
\mathbb{E}\Bigg[
\frac{1}{HW}
\sum_{u}
\mathrm{KL}\big(p(u)\,\|\,q(u)\big)
\Bigg], \\[3pt]
\mathcal{L}_{\mathrm{conf}} &=
\mathbb{E}\Bigg[
\frac{1}{HW}
\sum_{u}
(H(p(u)) + H(q(u)) - H\!(\frac{p(u)+q(u)}{2}))
\Bigg].
\end{aligned}
\]
}
Here, $u=(h,w)$ denotes a spatial pixel index, while $p(u)$ and $q(u)$ represent the predicted categorical distributions at location $u$ from $f^1_{\theta}$ and $f^2_{\phi}$, respectively. $\mathrm{KL}(\cdot)$ denotes the Kullback--Leibler divergence, and $H(\cdot)$ represents the Shannon entropy of a probability distribution. The entropy-based agreement term acts as a regularizer that encourages both models to produce consistent and confident predictions, thereby improving the robustness of \ac{SSL} with limited annotations.

\subsection{The Overall Objective Function}
\label{ssec:learningloss}

The FM-DACL framework is optimized using a joint objective that combines supervised learning and multiple semi-supervised regularization terms. Specifically, the overall loss consists of four components: a supervised loss $\mathcal{L}_{\text{sup}}$, a cross-supervision loss $\mathcal{L}_{\text{cps}}$, an interpolation consistency loss $\mathcal{L}_{\text{ict}}$, and a dual agreement consistency loss $\mathcal{L}_{\text{dac}}$. The total objective is defined as
\[
\label{loss}
\mathcal{L}_{\text{total}}
=
(\mathcal{L}_{\text{sup}}^{1}+\mathcal{L}_{\text{sup}}^{2})
+ \lambda(\mathcal{L}_{\text{cps}}^{1}+\mathcal{L}_{\text{cps}}^{2})
+ \tau \mathcal{L}_{\text{ict}}
+ \beta \mathcal{L}_{\text{dac}},
\]
where $\lambda$, $\tau$, and $\beta$ control the relative importance of
the semi-supervised losses.

The supervised losses $\mathcal{L}_{\text{sup}}^{1}$ and $\mathcal{L}_{\text{sup}}^{2}$ are computed on labeled samples $\boldsymbol{X}_{\text{L}}$ for the foundation model $f^1_{\theta}$ and the U-Net $f^2_{\phi}$, respectively. Each supervised loss
includes segmentation and classification terms:
{\small
\[
\begin{aligned}
\mathcal{L}_{\text{sup}}^{1} &= 
\CE(f^1_{\theta}(\boldsymbol{X}_{\text{L}}), \boldsymbol{Y}_{\text{gt}})
+ \Dice(f^1_{\theta}(\boldsymbol{X}_{\text{L}}), \boldsymbol{Y}_{\text{gt}})
+ \BCE(g^1_{\theta}(\boldsymbol{X}_{\text{L}}), \boldsymbol{C}_{\text{gt}}), \\
\mathcal{L}_{\text{sup}}^{2} &= 
\CE(f^2_{\phi}(\boldsymbol{X}_{\text{L}}), \boldsymbol{Y}_{\text{gt}})
+ \Dice(f^2_{\phi}(\boldsymbol{X}_{\text{L}}), \boldsymbol{Y}_{\text{gt}})
+ \BCE(g^2_{\phi}(\boldsymbol{X}_{\text{L}}), \boldsymbol{C}_{\text{gt}}),
\end{aligned}
\]
}
where $\CE(\cdot)$ and $\Dice(\cdot)$ denote the cross-entropy and Dice losses for segmentation, and $\BCE(\cdot)$ represents the binary cross-entropy with logits used for the multi-label classification task. The loss $\mathcal{L}_{\text{cps}}$ is defined in Eq.~(\ref{cps}), while $\mathcal{L}_{\text{ict}}$ and $\mathcal{L}_{\text{dac}}$ correspond to the interpolation consistency and dual agreement consistency terms described in the previous sections.

\section{Experiments and Results}

\subsection{Datasets}
The FETUS 2026 challenge provides a large-scale fetal echocardiography dataset for cardiac structure segmentation and CHD diagnosis. The dataset contains 5,000 standard-view B-mode ultrasound images collected from multiple clinical centers and ultrasound systems, including GE Voluson E10 and Philips Affiniti 70, capturing real-world characteristics such as motion artifacts, speckle noise, and variations in gestational age. The dataset also includes a subset of images derived from the publicly available FOCUS dataset \cite{Wu:2025}. Images cover four standard cardiac views: 4CH, LVOT, RVOT, and 3VT. Expert annotations provide segmentation masks for 14 anatomical structures and classification labels for seven CHD categories. Following the official split, 2,800 images are used for training, 600 for validation and 1,600 for testing. Among the training images, 20\% are labeled while the remaining 80\% are unlabeled.

\subsection{Implementation Details}

We evaluate \ac{FM-DACL} with different backbone combinations. The U-Net model is randomly initialized and trained from scratch. ResUNet uses a ResNet50 encoder initialized with ImageNet pretrained weights. SegFormer is initialized from publicly available pretrained weights, while EchoCare uses the official foundation model checkpoint provided by the authors~\cite{zhang:2025}. This setup allows evaluation of FM-DACL across both pretrained and non-pretrained architectures.

EchoCare is optimized using the AdamW optimizer with separate learning rates for different parameter groups. Specifically, the backbone parameters are trained with a learning rate of $10^{-4}$, while the task-specific heads use a higher learning rate of $10^{-3}$. The weight decay for AdamW is set to $0.01$. SegFormer and ResUNet are optimized using AdamW with a learning rate of $10^{-3}$ and weight decay of $10^{-4}$. The \ac{FM-DACL} framework is trained for 300 epochs with a labeled batch size of 1 and an unlabeled batch size of 4.
All experiments are implemented in Python (3.11.5) using PyTorch (2.1.2) with CUDA (12.2) and are conducted on a single NVIDIA Tesla GPU. 
The input images are resized to $256 \times 256$ pixels. Data augmentation includes random horizontal flipping and random rotation ($\pm20^\circ$). Loss weights are set to $\lambda= 5.0$, $\tau=1.0$ and $\beta=5.0$. 
Model performance is evaluated on the validation set after each training epoch, and the best-performing EchoCare checkpoint is selected based on the validation metrics.

\subsection{Evaluation Metrics}

Model performance is evaluated using the official challenge evaluation metrics. For the segmentation task, we report the Dice Similarity Coefficient (DSC) and Normalized Surface Dice (NSD). For CHD classification, we report the F1-score, which balances precision and recall and is particularly suitable for imbalanced multi-label classification problems. 
The overall challenge score is computed as:
\[
S_{score} = 0.45 \times S_{cls} + 0.45 \times S_{seg} + 0.1 \times S_{time}
\]
where the segmentation score is defined as
$
S_{seg} = \frac{\text{DSC} + \text{NSD}}{2}
$.
All metrics are computed on the validation set following the official challenge evaluation protocol.

\subsection{Results}

Table~\ref{tab:results} presents the performance of \ac{FM-DACL} with different heterogeneous backbone combinations. Due to limited development time during the challenge period, extensive hyperparameter tuning and architecture search were not performed; therefore, these experiments aim to demonstrate the feasibility and flexibility of the proposed dual-agreement framework rather than to optimize the leaderboard ranking. Among the tested configurations, the EchoCare--ResUNet combination achieves the best performance (DSC: 59.66, NSD: 42.82, Overall Score: 36.84), suggesting that combining pretrained heterogeneous backbones provides stronger feature complementarity. Although \ac{FM-DACL} does not outperform the official baseline, it demonstrates stable performance across different backbone choices, indicating that the framework is largely architecture-agnostic. These results suggest that further improvements may be achieved through systematic tuning and stronger backbone selection, while the main contribution of this work lies in the proposed \ac{SSL} framework rather than backbone optimization.

\begin{table}
\centering
\footnotesize
\setlength{\tabcolsep}{4pt}
\caption{Performance of FM-DACL with different backbone combinations on the validation set. $\star$: trained from scratch. $\dagger$: ImageNet-pretrained encoder.}
\begin{tabular}{lccccc}
\hline
Method ($f_1$ / $f_2$) & DSC $\uparrow$ & NSD $\uparrow$ & F1 $\uparrow$ & Score $\uparrow$ \\
\hline
Challenge Baseline & 65.48 & 45.55 & 34.20 & 40.37 \\
FM-DACL (EchoCare / U-Net$\star$) & 42.90 & 28.59 & 31.76 & 30.38 \\
FM-DACL (SegFormer$\dagger$ / ResUNet$\dagger$) & 45.80 & 30.22 & 37.35 & 33.91 \\
FM-DACL (EchoCare / ResUNet$\dagger$) & 59.66 & 42.82 & 30.62 & 36.84 \\
\hline
\end{tabular}
\label{tab:results}
\end{table}

\section{Conclusion}
In this work, we presented \ac{FM-DACL}, a semi-supervised framework for fetal heart ultrasound segmentation and CHD classification in the FETUS 2026 challenge. The proposed method combines a foundation model and convolutional networks through heterogeneous co-training with cross-supervision and dual agreement consistency to better exploit unlabeled data. Experiments demonstrate that the framework achieves stable performance across different backbone combinations. These results suggest that \ac{FM-DACL} provides a flexible \ac{SSL} strategy for fetal cardiac ultrasound analysis under limited annotation settings. Future work will focus on improving backbone selection and optimization strategies to further enhance performance.



\bibliographystyle{IEEEbib}
\bibliography{strings,refs}

\end{document}